# Energy-efficient field-free unconventional spin-orbit torque magnetization switching dynamics in van der Waals heterostructures

Lalit Pandey[1,2], Bing Zhao[1], Karma Tenzin[3], Roselle Ngaloy[1], Veronika Lamparská[3], Himanshu Bangar[1], Aya Ali[4], Mahmoud Abdel-Hafiez[5,6], Gaojie Zhang[7], Hao Wu[7], Haixin Chang[7], Lars Sjöström[1], Prasanna Rout[1], Jagoda Sławińska[3], Saroj P. Dash[1,2,8*]

[1]*Department of Microtechnology and Nanoscience, Chalmers University of Technology, SE-41296, Göteborg, Sweden.*
[2]*Wallenberg Initiative Materials Science for Sustainability, Department of Microtechnology and Nanoscience, Chalmers University of Technology, SE-41296, Göteborg, Sweden.*
[3]*Zernike Institute for Advanced Materials, University of Groningen, Nijenborgh 3, 9747AG Groningen, The Netherlands.*
[4]*Center for Advanced Materials Research, Research Institute of Sciences and Engineering, University of Sharjah, Sharjah 27272, United Arab Emirates*
[5]*Department of Applied Physics and Astronomy, University of Sharjah, Sharjah, United Arab Emirates.*
[6]*Department of Physics and Astronomy, Uppsala University, Box 516, SE-751 20 Uppsala, Sweden.*
[7]*School of Materials Science and Engineering, Huazhong University of Science and Technology, 430074, Hubei, China.*
[8]*Graphene Center, Chalmers University of Technology, SE-41296, Göteborg, Sweden.*

**Abstract**

Van der Waals (vdW) heterostructure of two-dimensional (2D) quantum materials offers a promising platform for efficient control of magnetization dynamics for non-volatile spin-based devices. However, energy-efficient field-free spin-orbit torque (SOT) switching and spin dynamics experiments to understand the basic SOT phenomena in all-2D vdW heterostructures are so far lacking. Here, we demonstrate energy-efficient field-free spin-orbit torque (SOT) switching and tunable magnetization dynamics in a vdW heterostructure comprising out-of-plane magnet $Fe_3GaTe_2$ and topological Weyl semimetal $TaIrTe_4$. We measured the non-linear second harmonic Hall signal in $TaIrTe_4$/$Fe_3GaTe_2$ devices to evaluate the SOT-induced magnetization dynamics, which is characterized by a large and tunable out-of-plane damping-like torque. Energy-efficient and deterministic field-free SOT magnetization switching is achieved at room temperature with a very low current density. First-principles calculations unveil the origin of the unconventional charge-spin conversion phenomena, considering the crystal symmetry and electronic structure of $TaIrTe_4$. These results establish that all-vdW heterostructures provide a promising route to energy-efficient, field-free, and tunable SOT-based spintronic devices.

Corresponding author: Saroj P. Dash, Email: saroj.dash@chalmers.se

## Introduction

In quantum materials, the interplay between spin-orbit coupling and magnetism, with additional control over the band topology, quantum geometries, and crystal symmetries can offer the potential for next-generation universal memory and computing technologies[1,2]. Specifically, enhanced functionalities can be achieved using efficient charge-spin conversion (CSC) phenomena in such quantum materials to enable spin-orbit torque (SOT) induced magnetization switching of a ferromagnet (FM)[3]. In conventional SOT memory devices, commonly used spin-orbit materials (SOM) exhibit moderate CSC efficiency and primarily generate in-plane SOT torque components, limiting their application in switching a magnet with perpendicular magnetic anisotropy (PMA)[4].

Recently developed van der Waals (vdW) heterostructures of two-dimensional (2D) SOMs and FMs can offer a new framework to address the challenges in SOT technologies[5]. Interestingly, low crystal symmetries of vdW SOMs can generate out-of-plane SOT components, making them suitable for field-free switching of ferromagnets with PMA[6–9]. Meanwhile, vdW magnets such as $Fe_3GeTe_2$ and $Fe_3GaTe_2$ with strong PMA are also developed, showing promises for reliable SOT device operations[10–13]. Taking advantage of such available quantum materials, all-2D vdW heterostructures have been explored for field-free SOT magnetization switching[14–17]. However, the SOT switching parameters are two to three orders of magnitude lower than required for energy-efficient switching.

To circumvent this issue, Weyl semimetals $TaIrTe_4$ with low crystal symmetry, large spin-orbit coupling (SOC), and large Berry curvature dipole were explored to generate a larger out-of-plane SOT component for energy-efficiency and field-free SOT switching of conventional magnets [8,9,18]. Therefore, all-2D vdW heterostructure combining the best vdW quantum materials with a large current-induced out-of-plane spin polarization and above room temperature vdW ferromagnet with an out-of-plane magnetization is a promising for energy-efficient non-volatile spintronic technologies. Furthermore, the investigation of magnetization dynamics in all-2D vdW heterostructures is critical for understanding the interplay between broken crystal symmetries, unconventional charge-spin conversion, and SOT-induced magnetization dynamics, ultimately enabling the design of efficient and ultrafast spintronic devices.

Here, we report tunable magnetization dynamics using harmonic measurements and demonstrate energy-efficient field-free SOT magnetization switching using the all-vdW heterostructures of $TaIrTe_4/Fe_3GaTe_2$ at room temperature. Weyl semimetal $TaIrTe_4$ with a tunable canted spin polarization combined with a vdW ferromagnet $Fe_3GaTe_2$ with strong PMA enable the exploration of magnetization dynamics and their tunable SOT efficiency. The 2[nd] harmonic measurements with detailed magnetic field and angle-dependent measurements at various temperatures reveal a large and tunable unconventional out-of-plane SOT torque in the $TaIrTe_4/Fe_3GaTe_2$ all-vdW heterostructure. The SOT components are observed to vary with temperature and correlate with the measured spin canting angle. Moreover, we observed a field-free deterministic SOT magnetization switching with a very low critical switching current density of $1.81 \times 10^{10}\ \mathrm{A/m^2}$, demonstrating energy-efficient non-volatile spintronic memory device. To unveil the origin of the unconventional CSC phenomena in $TaIrTe_4$, detailed first-principles calculations were performed considering crystal symmetry and electronic structures.

## Results and discussion

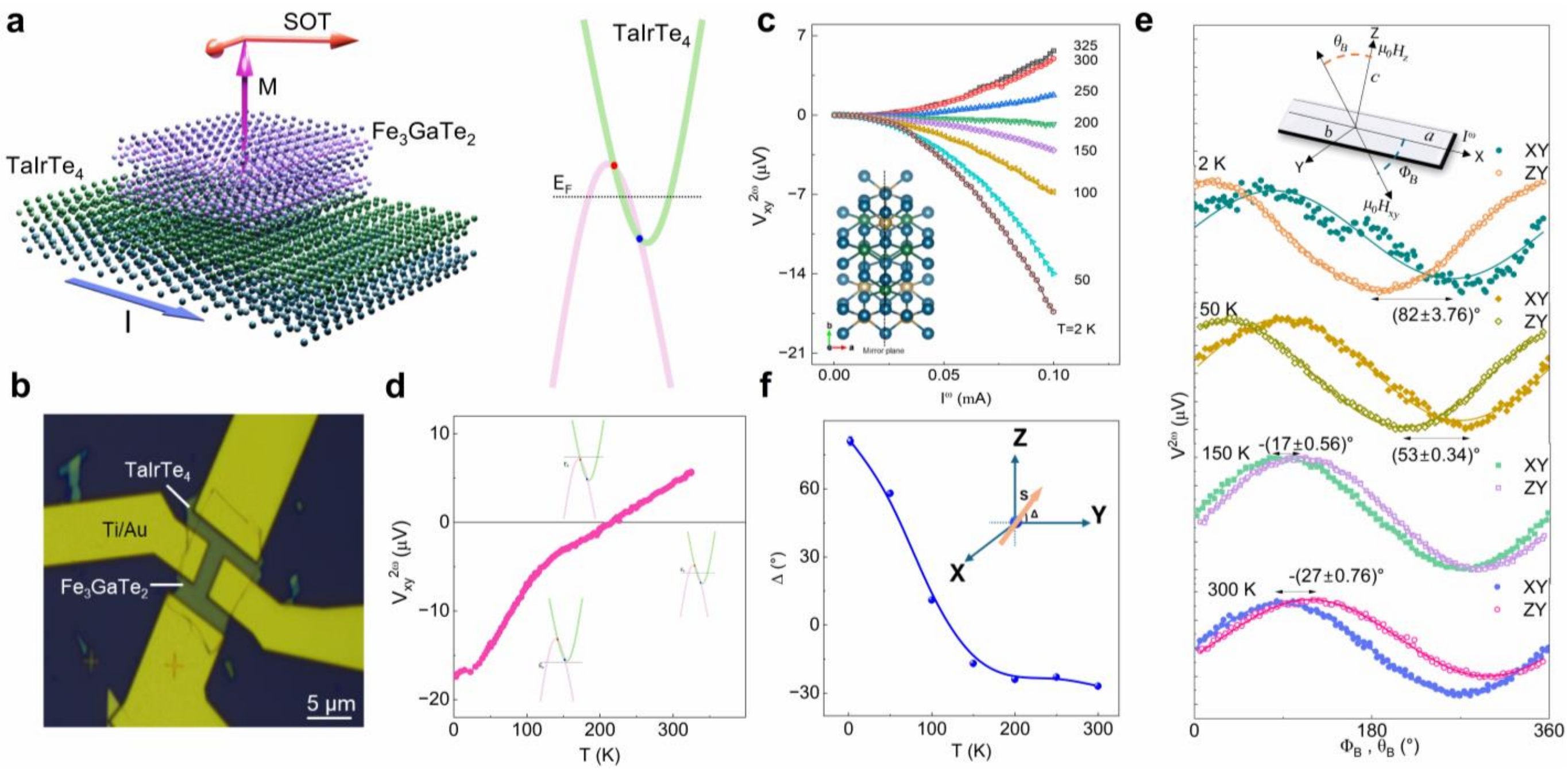

**Figure 1. Van der Waals heterostructure of $TaIrTe_4/Fe_3GaTe_2$ and harmonic measurements on $TaIrTe_4$. a.** Schematic diagram of a van der Waals heterostructure of Weyl semimetal $TaIrTe_4$ and out-of-plane ferromagnet $Fe_3GaTe_2$. Band structure of typical type-II Weyl semimetal with two Weyl nodes. **b.** Optical image of representative $TaIrTe_4/Fe_3GaTe_2$ vdW heterostructure Hall bar device with a scale bar of 5 μm. **c.** 2nd harmonic transverse Hall voltage $V_{xy}^{2\omega}$ in response to an applied alternating current $I^{\omega}$ along a-axis at different temperatures for a device with 20 nm $TaIrTe_4$. The inset illustrates the crystal structure of $T_d$-$TaIrTe_4$, characterized by low crystal symmetry and a mirror plane along the crystallographic b-axis. **d.** 2nd harmonic transverse Hall voltage $V_{xy}^{2\omega}$ with temperature at an $I^{\omega}$ of 0.1 mA of $TaIrTe_4$. Insets show the energy dispersion curve of type-II Weyl semimetal and tuning of Fermi level energy ($E_F$) with temperature. **e**. 2nd harmonic voltage $V^{2\omega}$ response measured in $TaIrTe_4$ device as a function of angle between current applied along a-axis of $TaIrTe_4$ ($|I^{\omega}|$ = 0.1 mA) and external magnetic field (13 T). The device is rotated in XY and ZY planes, as depicted in schematics. In the XY rotation, the device rotates such that the magnetic field aligns parallel to the sample surface and making $\Phi_B$ angle with a-axis of $TaIrTe_4$, whereas in ZY rotation, the device rotation changes magnetic field direction from a-axis of $TaIrTe_4$ to c axis and making $\theta_B$ angle with c-axis with $TaIrTe_4$. The solid lines are the fits. **f.** Temperature dependence of shift (Δ) in the maxima or minima of $V^{2\omega}$ vs $\Phi_B$ and $\theta_B$ curves. This shift is denoted as out-of-plane canting angles, as illustrated in schematics. Such shift is directly correlated to the out-of-plane spin canting angle, which is estimated to be − (27±0.76) ° at room temperature.

We investigated $TaIrTe_4/Fe_3GaTe_2$ vdW heterostructures (Fig. 1a) due to their promising properties, anticipating that their combination could yield new phenomena such as large non-linear Hall effects and unconventional spin-orbit torque (SOT) magnetization dynamics. $TaIrTe_4$ is a vdW topological Weyl semimetal (WSM) candidate, with a significant Berry curvature dipole and large spin splitting of the electronic bands[19]. In addition, it provides unconventional charge-spin conversion with an out-of-plane spin polarization component that can induce an out-of-plane SOT[18] on the adjacent PMA ferromagnet to induce a magnetic field-free switching. On the other hand, $Fe_3GaTe_2$ is a unique vdW topological nodal line metallic ferromagnet with strong PMA above room temperature with $T_c$ around 370 K[10]. We fabricated Hall-bar devices based on $TaIrTe_4/Fe_3GaTe_2$ vdW heterostructures, along with individual Hall bars on $TaIrTe_4$ and $Fe_3GaTe_2$ crystals, to characterize properties, such as the anomalous Hall effect (AHE), 2$^{nd}$ Harmonics measurements and SOT-driven switching experiments (details in Methods section and Supplementary Fig. S1). Figure 1b presents a typical optical microscope image of a representative $TaIrTe_4/Fe_3GaTe_2$ Hall-bar device.

**Tunable spin texture using bilinear magnetoresistance and 2$^{nd}$ harmonic Hall effect in $TaIrTe_4$**

$TaIrTe_4$ exhibited a strong nonlinear Hall effect, characterized by a 2$^{nd}$ harmonic Hall voltage that nonlinearly depends on driving currents sourced along the a-axis of the crystal, perpendicular to its mirror plane at room temperature (Fig. 1c). Unlike linear Hall effects observed in systems with broken time-reversal symmetry, the nonlinear Hall effect in $TaIrTe_4$ arises from the large Berry curvature dipole in the absence of inversion-symmetry. Notably, the nonlinear Hall voltage changed sign near ~200 K (Fig. 1c and 1d), indicating temperature-induced shift in the chemical potential, consistent with the Weyl semi metallic properties of $TaIrTe_4$[20]. The current induced spin polarization in $TaIrTe_4$ are probed using bilinear magnetoelectric resistance (BMER) technique[9,21], measuring 2$^{nd}$ harmonic voltage while rotating the samples in XY and ZY planes (Fig. 1e). In XY rotation, the magnetic field vector remains in the ab crystallographic plane sweeping azimuthal angle ($\Phi_B$) with respect to the a-axis of $TaIrTe_4$, whereas in ZY rotation, the field vector sweeps polar angle ($\theta_B$) with respect to the c-axis of $TaIrTe_4$ in the ac plane. Figure 1e depicts the temperature dependence of 2$^{nd}$ harmonic voltage with $\Phi_B$ and $\theta_B$. The direction of resultant spin angular momentum arises due to charge-spin conversion effects in $TaIrTe_4$ being equivalent to angular shift ($\Delta$) of BMER curves measured along XY and ZY geometries. The $\Delta$ is found to be $-(27 \pm 0.76)°$ at room temperature, indicating the presence of an out-of-plane spin density induced in $TaIrTe_4$. Such spin-polarization can help in generating unconventional out-of-plane SOT in adjacent ferromagnetic layer $Fe_3GaTe_2$ with PMA resulting in field-free deterministic switching. The temperature dependence of $\Delta$ (Fig. 1f) suggests that the polarity and magnitude of the spin canting angle in $TaIrTe_4$ are highly tunable by the position of chemical potential/Fermi level[9].

**Perpendicular magnetic anisotropy of $Fe_3GaTe_2$**

To verify the magnetic property and anisotropy of $Fe_3GaTe_2$, the anomalous Hall resistance $R_{xy}$ is measured at different temperatures ranging from 2 to 300 K (Fig. 2a, b). A square-shaped magnetic hysteresis loop is observed with coercivity around 100 mT and anomalous Hall resistance ($R_{AHE}$) of around 1.5 Ω at room temperature, where the latter is directly proportional to saturation magnetization ($M_s$) of $Fe_3GaTe_2$. The $R_{AHE}$ vs T curve, shown in Fig. 2c, is fitted with $R_{xy}(T) = R_{xy}(0)\left(1-\left(\frac{T}{T_c}\right)^2\right)^{\beta}$ analogues to Bloch equation for magnetization vs temperature curve to estimate Curie temperature ($T_c$ = $369.14 \pm 7.73$ K)[10] and critical magnetization exponent $\beta = 0.35$ [22–25]. Figure 2d shows the anomalous Hall resistance of $Fe_3GaTe_2$ as a function of in-plane magnetic fields at different temperatures from 2 to 300 K. A magnetic hysteresis loop is observed at all temperatures, with finite remanence and coercivity, consistent with the typical behavior of PMA magnets along their hard axis (Fig. 2e).

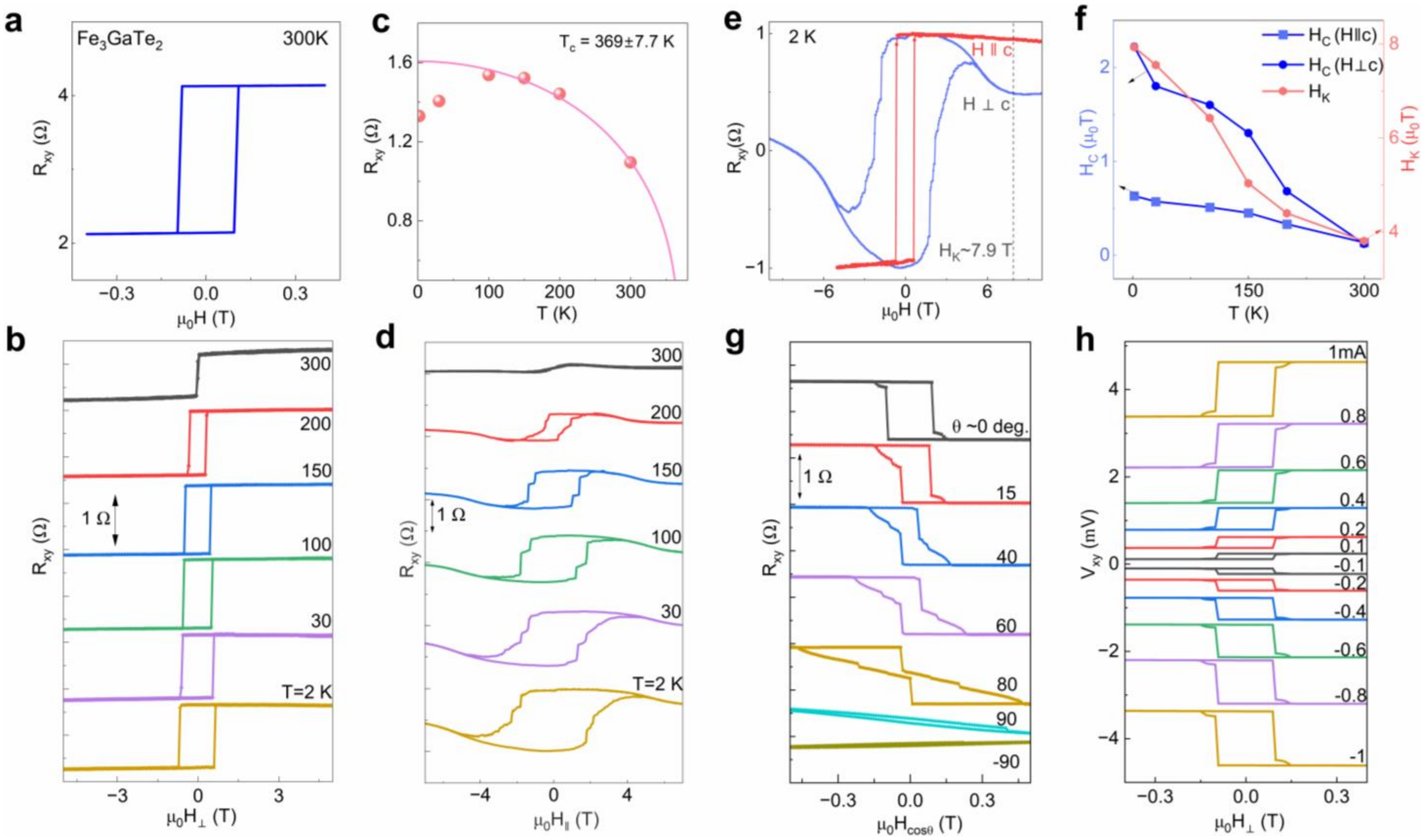

**Figure 2. Magneto-transport characterization of $Fe_3GaTe_2$. a,b,** Anomalous Hall resistance of $Fe_3GaTe_2$ as a function of out-of-plane magnetic fields at 300 K and temperature dependence ranging from 2 to 300 K. **c.** Anomalous Hall amplitude at the saturated field as a function of temperature. Solid line is fit to extract the Curie temperature ($T_c = 369.14 \pm 7.73$ K) **d.** Anomalous Hall resistance of $Fe_3GaTe_2$ as a function of in-plane magnetic fields at different temperatures ranging from 2 to 300 K. **e.** Comparison of anomalous Hall effect measurement for field swept parallel to sample plane (i.e., $H \perp c$) vs perpendicular (i.e., $H \parallel c$) to sample plane at 2 K temperature. The anisotropic field ($H_K$) is ~7.9 T, indicating strong perpendicular magnetic anisotropy present in $Fe_3GaTe_2$. **f.** Variation of coercive fields and anisotropic fields with temperature extracted from ($R_{xy}$ vs $\mu_0 H_{\perp}$) and ($R_{xy}$ vs $\mu_0 H_{\parallel}$) measurements. **g.** AHE signals $R_{xy}$ with different out-of-plane angles (θ) between the magnetic field and the c-axis of the sample plane at 300 K. **h.** Variation of AHE signals $R_{xy}$ with positive and negative DC bias currents.

Figure 2f shows the variation of magnetic coercivity ($H_c$) in both field directions (i.e., $H \perp c - axis$ and $H \parallel c - axis$) and anisotropic field with temperature. The anisotropic field ($H_K$), defined as the difference in saturation between in-plane and out-of-plane magnetic fields, reaches ~7.9 T at 2 K and ~3.8 T at 300 K. Such a high value of $H_K$ suggests that $Fe_3GaTe_2$ has a very high magnetic anisotropy energy density with a very strong PMA. The coercive field ($H_c$) is also quite high along in-plane direction as compared to out-of-plane direction. Both the $H_c$ and $H_K$ decrease with an increase in temperature approaching the Curie temperature of $Fe_3GaTe_2$. Figure 2g illustrates AHE signals $R_{xy}$ measured at varying out-of-plane angles (θ) between c-axis of sample and magnetic field. It can be noted here that the magnitude of AHE signal ($R_{xy}^{AHE} = \frac{R_{xy}(+H_S) - R_{xy}(-H_S)}{2}$) remains almost constant till $\pm 80^o$; beyond that AHE loop disappears between $\pm 600$ mT field range. Again, this indicates a strong out-of-plane magnetic anisotropy present in $Fe_3GaTe_2$. Figure 2h shows the variation of AHE signals $R_{xy}$ with positive and negative DC bias currents. We observed that the magnitude of anomalous Hall signal, the coercivity and saturation fields remain unchanged with positive or negative current bias varied to $\pm 0.1$ mA to $\pm 1$ mA, indicating the robustness of perpendicular anisotropic magnetic moment against dc current within these bias ranges.

### 2nd harmonic nonlinear Hall effect and spin-orbit torque induced magnetization dynamics in $TaIrTe_4/Fe_3GaTe_2$ heterostructures

The harmonic Hall measurements are performed on $TaIrTe_4/Fe_3GaTe_2$ heterostructures to quantitatively evaluate the non-linear effects and magnetization dynamics driven by SOT. When a sinusoidal current ($I^{\omega}$) is applied to the vdW heterostructure, composed of the spin-orbit material $TaIrTe_4$ and a ferromagnet $Fe_3GaTe_2$, spin-orbit torques ($\tau_{SOT}$) are exerted on the magnetization (m) of the $Fe_3GaTe_2$. This effect originates from the spin accumulation at the vdW interface due to efficient charge-spin conversion in $TaIrTe_4$. Typically, two mutually orthogonal torques are generated: the damping-like torque ($\tau_{DL} \sim m \times (\sigma \times m)$) and the field-like torque ($\tau_{FL} \sim \sigma \times m$ )[11,26].

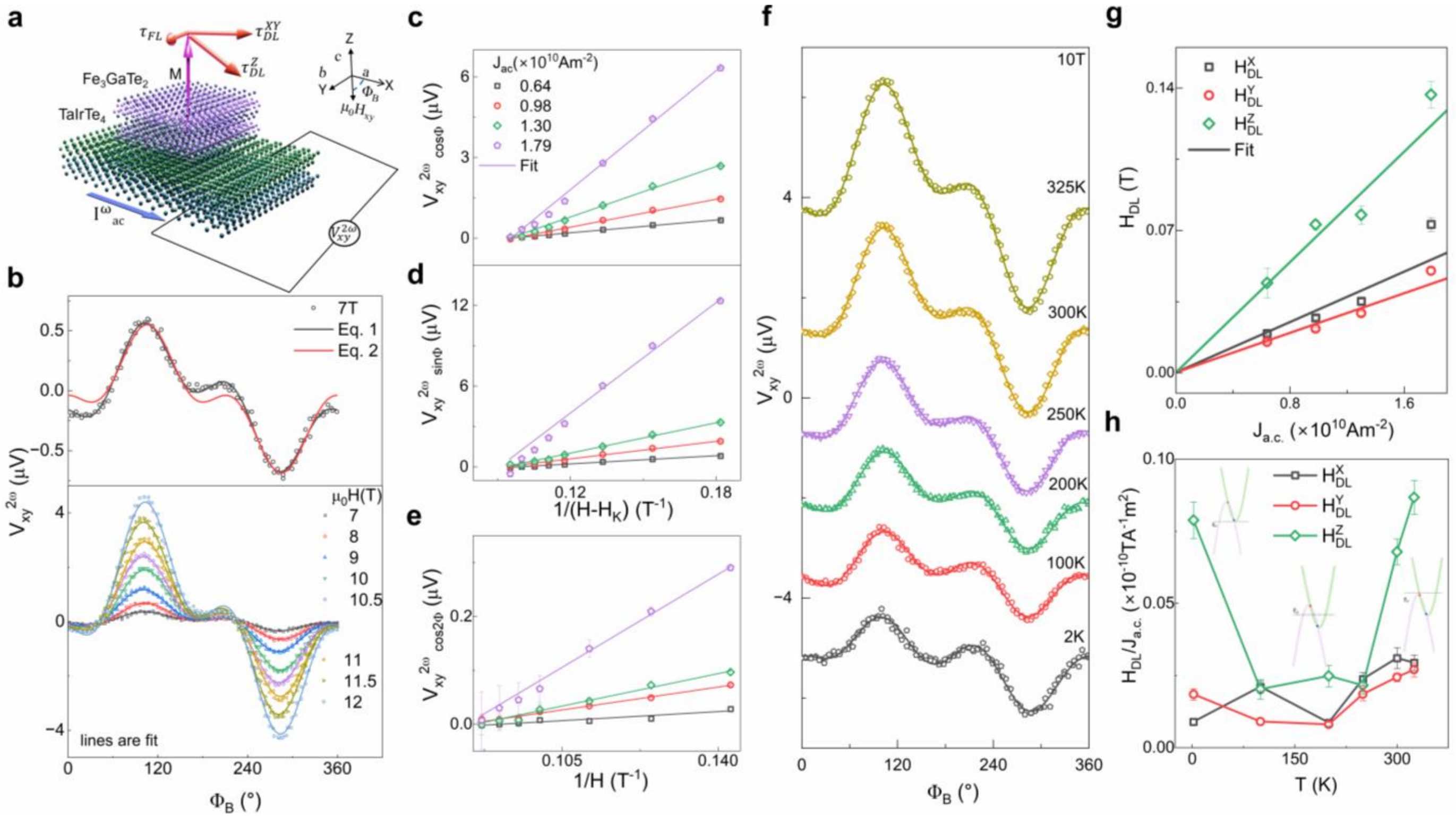

**Figure 3**. **Angle-dependent harmonic Hall measurements in $TaIrTe_4/Fe_3GaTe_2$ heterostructure. a.** Schematic of the $TaIrTe_4/Fe_3GaTe_2$ heterostructure, illustrating the effects of damping-like torques ($\tau_{DL}^{XY}$ and $\tau_{DL}^{Z}$) and field-like torques ($\tau_{FL}$) on $Fe_3GaTe_2$ magnetization when the current is applied along the a-axis of $TaIrTe_4$ layer. The 2nd harmonics Hall voltage ($V_{xy}^{2\omega}$) measurement scheme is shown with an external in-plane magnetic field at angle $\Phi_B$ relative to the a.c. current direction $I_{ac}$. **b.** $V_{xy}^{2\omega}$ vs $\Phi_B$ of Dev1 at magnetic field 7 T and temperature 300 K. The solid lines are fitted with Eq. 1 and 2. The second panel shows $V_{xy}^{2\omega}$ vs $\Phi_B$ for varied magnetic fields (7-12 T). **c,d,e.** Coefficient $V_{xy}^{2\omega}{}_{\cos\Phi_B}$($cos\Phi_B$ dependent in $V_{xy}^{2\omega}$), $V_{xy}^{2\omega}{}_{\sin\Phi_B}$($sin\Phi_B$ dependent in $V_{xy}^{2\omega}$) and $V_{xy}^{2\omega}{}_{\cos2\Phi_B}$($cos2\Phi_B$ dependent in $V_{xy}^{2\omega}$) as a function of 1/(H-$H_k$) and 1/H under different current densities $J_{a.c.}$. **f.** Angle sweep of $V_{xy}^{2\omega}$ at different temperatures (2-325 K) at a constant magnetic field of 10 T. Solid lines are fit to experimental data using equation 1. **g.** Damping-like field components ($H_{DL}^{X}, H_{DL}^{Y}, H_{DL}^{Z}$) as a function of current density, with linear fits estimating $H_{DL}/J_{a.c.}$. **h.** Temperature dependence of $H_{DL}/J_{a.c.}$ for $TaIrTe_4/Fe_3GaTe_2$ device. Insets show the energy dispersion curve of type-II Weyl semimetal and tuning of Fermi level energy ($E_F$) with temperature.

In these measurements, applying a sinusoidal current ($I^{\omega}$) with a fixed frequency of 213.3 Hz induces SOT-driven magnetization oscillation, generating harmonics in both the longitudinal and transverse resistance signals. The 1st and 2nd harmonic signals are measured and analyzed across various angles ($\Phi_B$) between the in-plane magnetic field ($H \perp c$) and the applied sinusoidal current ($I^{\omega}$), as well as under varying external magnetic fields ($H_{ext}$). This analysis provides information about the current-induced effective SOT fields and torques.

Since the spin Hall effect (SHE) in $TaIrTe_4$ induces both in-plane and out-of-plane spin polarizations ($\sigma^{X,Y,Z}$), the applied $I^{\omega}$ along the a-axis of $TaIrTe_4$ generates corresponding components of the damping-like ($\tau_{DL}^{X,Y,Z}$) and field-like ($\tau_{FL}^{X,Y,Z}$) torques. The 2nd harmonic transverse voltage generated from these current-induced effective SOT fields ($H_{DL}^{X,Y,Z}, H_{FL}^{X,Y,Z}$) and torques ($\tau_{DL}^{X,Y,Z}, \tau_{FL}^{X,Y,Z}$) in PMA ferromagnets is expressed as[27,28],

$$V_{xy}^{2\omega} = V_{DL}^{Y} cos\Phi_B + V_{DL}^{X} sin\Phi_B + V_{DL}^{Z} cos2\Phi_B + V_{FL}^{Y} cos\Phi_B cos2\Phi_B + V_{FL}^{X} sin\Phi_B cos2\Phi_B + V_{FL}^{Z} \quad \text{(Eq. 1)}$$

Here, damping-like torque components generated by X, Y and Z spin polarization contribute to coefficients $V_{DL}^{X,Y,Z}$, and the field-like torque counterparts give rise to coefficients $V_{FL}^{X,Y,Z}$. The estimation of SOT fields ($H_{DL}^{X,Y,Z}, H_{FL}^{X,Y,Z}$) from coefficients ($V_{DL}^{X,Y,Z}, V_{FL}^{X,Y,Z}$) is detailed in Supplementary Note 6 (Eqs. S1-S7).

The 2nd harmonic Hall signal as a function of $\Phi_B$ at constant fields ($H>H_K$) is plotted in Fig. 3b. The $V_{xy}^{2\omega}$ vs $\Phi_B$ curve is fitted with Eq. 2 ($V_{xy}^{2\omega} = V_{DL} cos\Phi_B + \mathrm{V_{FL}} cos\Phi_B cos2\Phi_B$) to estimate SOT in materials containing only conventional in-plane spins[29,30]. However, $V_{xy}^{2\omega}$ vs $\Phi_B$ curve of $TaIrTe_4/Fe_3GaTe_2$ could not be well fitted with Eq. 2 (see Fig. 3b). For proper fitting, we need to include $V_{DL} cos2\Phi_B$ and $V_{DL} sin\Phi_B$ terms as in Eq. 1, which consider additional torque components due to current-induced out-of-plane spin canting in $TaIrTe_4$. The coefficients ${V_{xy}^{2\omega}}_{\cos\Phi_B}$, ${V_{xy}^{2\omega}}_{\sin\Phi_B}$ and ${V_{xy}^{2\omega}}_{\cos 2\Phi_B}$ are hyperbolic functions of the magnetic field (see Supplementary Note 6, Eqs S2-S4), implying linear function on $1/(H-H_K)$ or $1/H$. The values of these coefficients were extracted by fitting the experimental 2nd harmonic transverse voltage and plotted in Fig. 3c-e. From these slopes, $H_{DL}^{X,Y,Z}$ are estimated (using $R_{AHE}$=1 Ω and $R_{PHE}$=0.0113 Ω; see Supplementary Note 5) and plotted as a function of current densities $J_{a.c.}$ in Fig. 3g. The slope of $H_{DL}^{X,Y,Z}$ vs $J_{a.c.}$ are found out to be $H_{DL}^{X}/J \sim (3.09 \pm 0.37) \times 10^{-12}\ \mathrm{TA^{-1}m^2}$, $H_{DL}^{Y}/J \sim (2.43 \pm 0.15) \times 10^{-12}\ \mathrm{TA^{-1}m^2}$ and $H_{DL}^{Z}/J \sim (6.78 \pm 0.44) \times 10^{-12}\ \mathrm{TA^{-1}m^2}$. This analysis indicates that current-induced effective SOT fields or torques in $TaIrTe_4/Fe_3GaTe_2$ heterostructure originated from the out-of-plane spin polarization ($\sigma^{Z}$), and it is larger than its in-plane counterparts ($\sigma^{XY}$).

**Tunable spin-orbit torque with temperature due to Fermi level tuning of $TaIrTe_4$**

The polarity and magnitude of spin accumulation generated by $TaIrTe_4$ are influenced by the chemical potential[9], resulting in temperature dependence changes in the tilt angle (as shown in Fig. 1f). A similar trend is expected in the current-induced effective SOT fields or torques. Hence, to observe the temperature dependence of SOT efficiency from the 2nd harmonic Hall signal of $TaIrTe_4/Fe_3GaTe_2$, angle sweep SHH measurements are conducted at different temperatures. Figure 3f illustrates the $V_{xy}^{2\omega}$ vs $\Phi_B$ curve at 10 T across different temperatures. Damping-like SOT effective fields ($H_{DL}^{X,Y,Z}$) are estimated using Eq. 1. $H_{DL}^{Z}/J_{a.c.}$ is highly tunable with temperature (Fig. 3h), it decreased from 2 K to 100 K, reached a minimum between 100-200 K, and increased from 200 K to 325 K. This behavior aligns with the temperature dependence of current-induced spin accumulation (Fig. 1f), showing large out-of-plane spin polarization at 2 K and room temperature with a minimum near 100 K. Hence, the out-of-plane damping-like torque is observed to be tunable by the chemical potential of $TaIrTe_4$.

**Field-dependent harmonic Hall measurements**

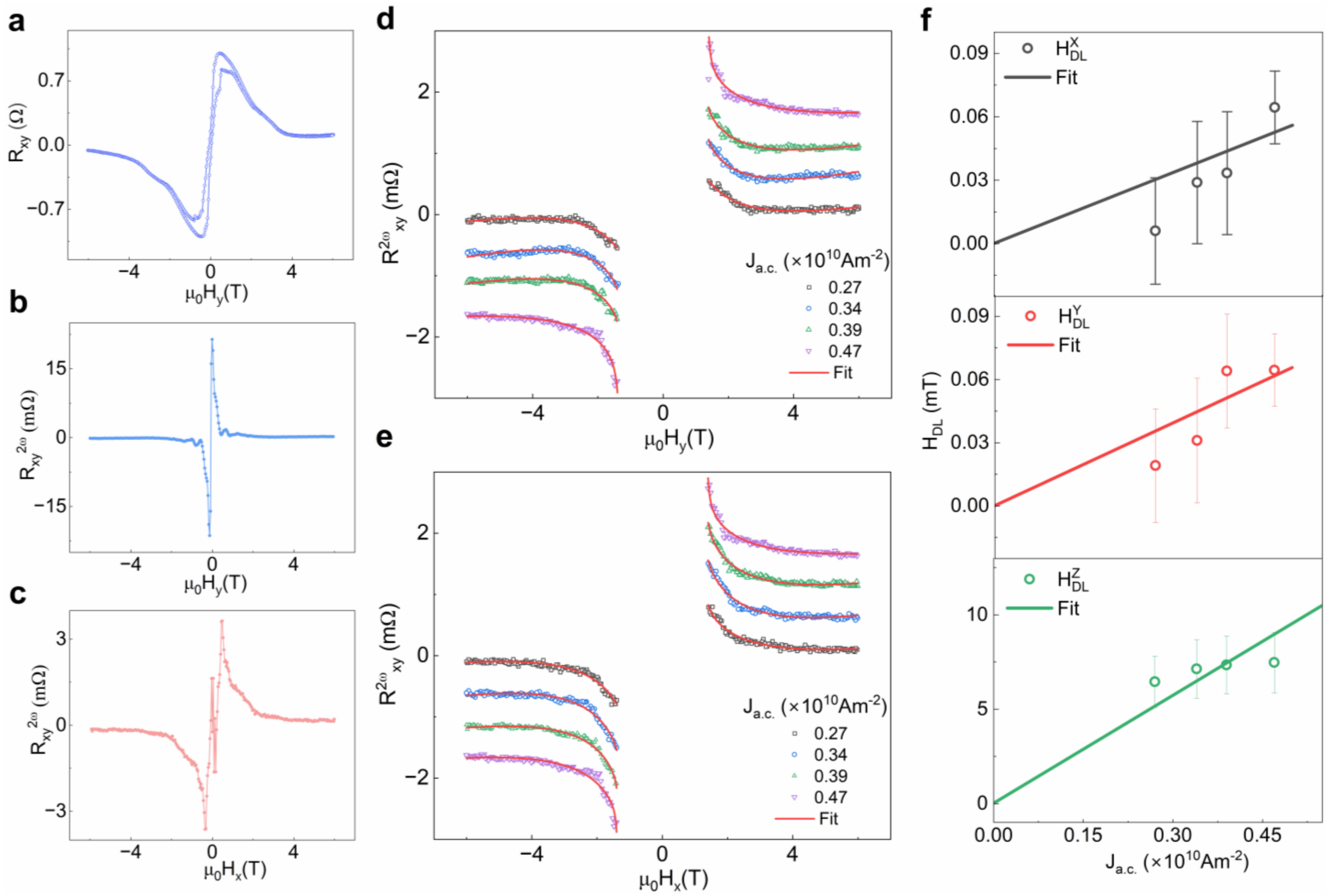

**Figure 4. Field-dependent harmonic Hall measurements in $TaIrTe_4/Fe_3GaTe_2$ heterostructure. a.** 1st harmonic transverse resistance ($R_{xy}^{1\omega}$) as a function of magnetic field swept parallel to the sample surface ($H\perp c$) and perpendicular to current direction, measured at 300 K on Dev 2. **b,c,** 2nd harmonic transverse resistance $R_{xy}^{2\omega}$ varied as a function of the external magnetic field applied along parallel to the sample surface, with $H_y$ representing $H\perp c$ and perpendicular to the current ($H\perp J$ ), and $H_x$ representing $H\perp c$ and parallel to the current ($H \parallel J$). **d,e,** Dependence of the 2nd harmonic transverse resistance ($R_{xy}^{2\omega}$) on the in-plane magnetic field ($H_y$ and $H_x$) for different magnitudes of constant write current density ($J_{a.c.}$). **f.** Extracted effective damping-like field components ($H_{DL}^{X,Y,Z}$) corresponding to spin polarization ($\sigma^X, \sigma^Y, \sigma^Z$ ) as a function of $J_{a.c.}$, obtained from fits to the 2nd harmonic signal.

To further validate and estimate SOT components, we measured the 1st and 2nd harmonic transverse Hall resistance $R_{xy}$ signal as a function of magnetic field applied parallel to the sample surface ($H\perp c$) and perpendicular to the applied current direction. In the 1st harmonic $R_{xy}^{\omega}$ vs H, a hysteresis loop with a magnetic anisotropic field $H_K$ of ~ 1.5 T is observed (Fig. 4a). The 2nd harmonics transverse Hall resistance signal $R_{xy}^{2\omega}$ varied with the external magnetic field applied parallel to the sample surface. The measurements are conducted with the field oriented either perpendicular ($H_y, \Phi_B = 90°$ ) or parallel ($H_x, \Phi_B = 0°$) to the direction of current (or a-axis of $TaIrTe_4$). These results are displayed in Figs. 4b and 4c. The resistance exhibited a hyperbolic dependence on the field for $|H|>H_k$, however became discontinuous for $|H| < H_K$.

Figures 4d and 4e show 2nd harmonics transverse resistance ($R_{xy}^{2\omega}$) versus $H_y$ and $H_x$ for different applied sinusoidal current densities ($J_{a.c.}$). The hyperbolic curvature of these plots sharpens with increasing current density. For $R_{xy}^{2\omega}$ vs $H_y$ data at $\Phi_B = 90°$, equation (1) reveals that only x and z components of the SOT fields contributing to the 2nd harmonics signal. Thus equation (1) simplifies to:

$$R_{xy}^{2\omega} = H_{DL}^{X}\frac{R_{AHE}}{2(H_y-H_K)} + H_{DL}^{Z}\frac{R_{PHE}}{H_y} + H_{FL}^{X}\frac{R_{PHE}}{H_y} + H_{FL}^{Z}\frac{R_{AHE}}{2(H_y-H_K)} \quad \text{(Eq. 3)}$$

From the analysis of second harmonic angular dependence data, we have concluded that $H_{FL}^{X}$ and $H_{FL}^{Z}$ make negligible contributions to the fitting of $V_{xy}^{2\omega}$ vs $\Phi_B$ curve. This reduces the fitting parameter, allowing precise calculation of $H_{DL}^{X}$ and $H_{DL}^{Z}$. Similarly, for $R_{xy}^{2\omega}$ vs $H_x$ data ($\Phi_B = 0°$), the extracted $H_{DL}^{Z}$ values are used to determine $H_{DL}^{Y}$ and $H_{FL}^{Y}$ (see Eq. 1 and Supplementary Note 6). The extracted values of $H_{DL}^{X,Y,Z}$ with different current densities $J_{a.c.}$ are plotted in Fig. 4f. The slopes of $H_{DL}$ vs $J_{a.c.}$ are found to be: $H_{DL}^{X}/J_{a.c.}$~ $(0.012 \pm 0.002) \times 10^{-12}\ \mathrm{TA^{-1}m^2}$, $H_{DL}^{Y}/J_{a.c.}$~ $(0.013 \pm 0.0015) \times 10^{-12}\ \mathrm{TA^{-1}m^2}$ and $H_{FL}^{Z}/J_{a.c.}$~ $(1.905 \pm 0.17) \times 10^{-12}\ \mathrm{TA^{-1}m^2}$. These findings also confirm that effective damping like field corresponding to Z spin polarization is significantly larger than those from XY polarized spins.

It should be noted that $TaIrTe_4$ alone also exhibits a 2nd harmonic voltage signal as function of $\Phi_B$, arising from broken mirror symmetry and finite Berry curvature dipole. This signal follows a $cos\Phi_B$ or $sin\Phi_B$ dependence (Fig. 1e). So, the $H_{DL}^{X}$ and $H_{DL}^{Y}$ values from the fitting of $V_{xy}^{2\omega}$ vs $\Phi_B$ data can be overestimated (Eq. 1). However, the estimation of Z-component damping-like field $H_{FL}^{Z}$ using coefficient $V_{xy}^{2\omega}{}_{\cos 2\Phi_B}$ (Eq. 1), central to the conclusion of second harmonic measurements, remains consistent. Furthermore, only $TaIrTe_4$ also has a unique field dependence of 2nd harmonics signal as shown in Supplementary Fig. S5c, which is quite different from SOT-induced $R_{xy}^{2\omega}$ vs H curve (see Fig. 4b-e). At large magnetic field, the $R_{xy}^{2\omega}$ vs H curve from $TaIrTe_4$ appears to be linear function of magnetic field; hence, to account for the 2nd harmonics field contribution of $TaIrTe_4$ and thermal effects[30,31], a linear polynomial term is included while fitting the field-dependent curves.

**Field-free deterministic spin-orbit torque switching in $TaIrTe_4$/$Fe_3GaTe_2$ heterostructure**

SOT magnetization switching experiments are crucial for investigating magnetization switching characteristics, such as determining the critical switching current density, assessing the need for an external field to aid in switching, and identifying whether the process is deterministic or non-deterministic. A series of pulse currents ($I_{pulse}$) applied along the a-axis in the $TaIrTe_4$/$Fe_3GaTe_2$ heterostructure can induce an unconventional spin current along the z-axis, with spin polarization $\sigma^Z$ oriented along the z-axis in $TaIrTe_4$[9]. This spin current generates an unconventional SOT on $Fe_3GaTe_2$, consisting of both field-like ($\tau_{FL}$) and damping-like ($\tau_{DL}$) torques, facilitating the switching of the magnetization direction M. The field-like torque $\tau_{FL} \sim \sigma \times m$ induces the precession of M around the exchange field generated by spin polarization, while the damping-like torque $\tau_{DL} \sim m \times (\sigma \times m)$ aligns M with the spin polarization $\sigma$, predominantly driving the magnetization switching (Fig. 4a)[32]. Figure 4b shows the AHE at 300 K of Dev3 used for switching experiments. Figure 4c presents SOT-induced magnetization switching, measured by applying a pulsed write current ($I_p$) along the a-axis with a pulse duration of 50 µs. This is followed by a small DC read current ($I_r$~500 µA) to determine the magnetization state via the Hall resistance $R_{xy}=V_{xy}/I_r$. Due to a large unconventional SOT, fully deterministic field-free magnetization switching could be observed at room temperature with $I_p$=±3.5 mA. Since the signal $R_{xy}$ is proportional to the out-of-plane magnetization $M_z$, the SOT $R_{xy}$ signal indicates a current-induced magnetization change between +M and -M. Notably, deterministic SOT switching of $TaIrTe_4/Fe_3GaTe_2$ heterostructure is observed at $H_x$=0 T, which indicates the creation of $\sigma^Z$ spin polarization in $TaIrTe_4$ leading to an out-of-plane SOT component. The magnitude of the switching signal is comparable to the AHE signal magnitude with field sweep, showing a full magnetization switching.

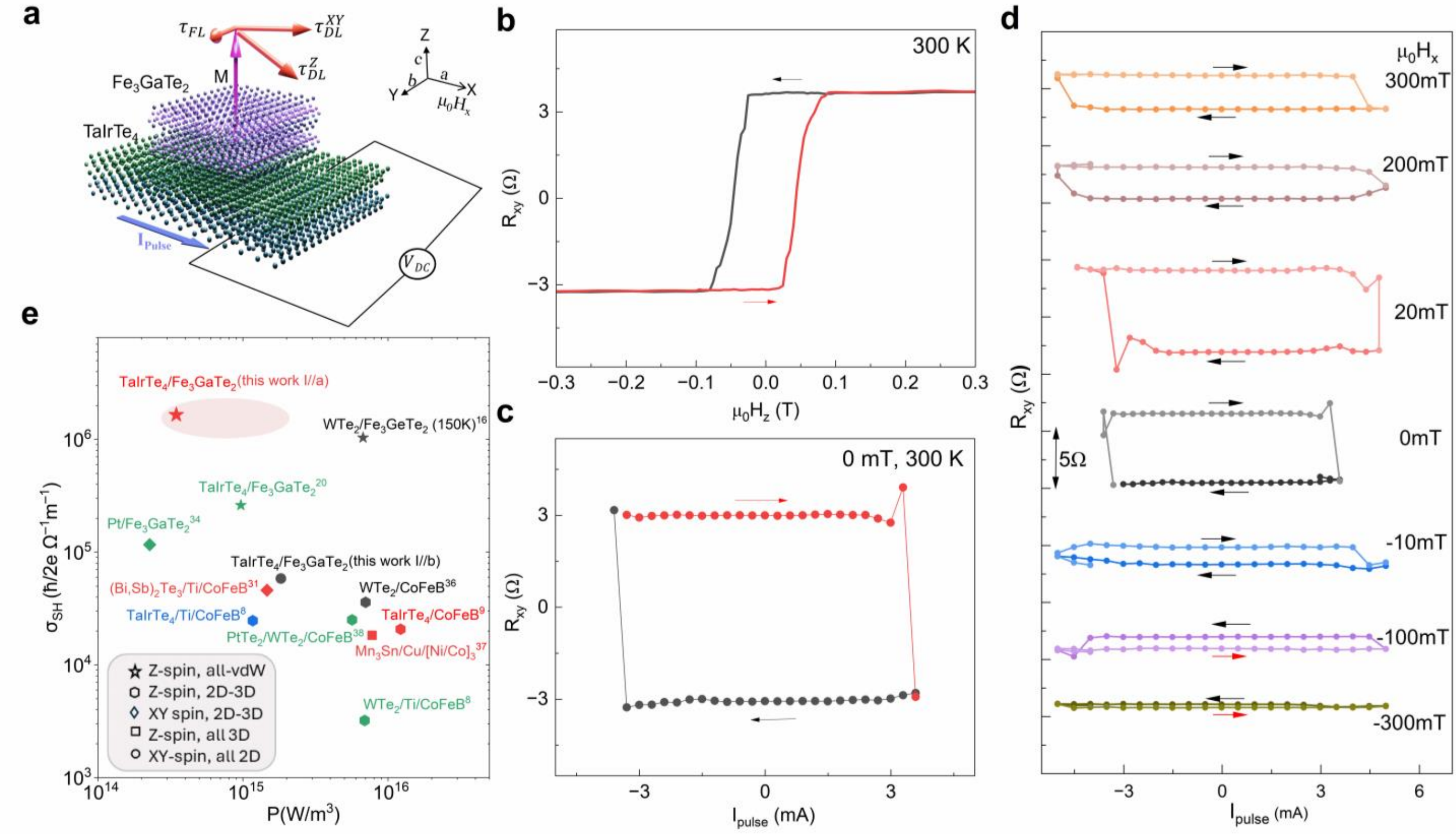

**Figure 5. Energy-efficient, field-free deterministic magnetization switching by spin-orbit torque in the $TaIrTe_4/Fe_3GaTe_2$ heterostructure at room temperature. a.** Diagrammatic representation of $TaIrTe_4/Fe_3GaTe_2$ heterostructure. This configuration leads to a significant out-of-plane antidamping torque ($\tau_{AD}^{OOP}$), which is symmetric with respect to the current direction, facilitating field-free deterministic switching of the $Fe_3GaTe_2$ magnetization. **b.** AHE of the $TaIrTe_4/Fe_3GaTe_2$ heterostructure device 3 with magnetic field sweep at 300 K. **c.** Field-free full deterministic switching achieved at 3.5 mA pulse current and 500 µA current is used as reading current to measure magnetization states keeping external field zero at 300 K temperature. The current is applied along the symmetry axis (a-axis) of $TaIrTe_4$. **d.** Current-driven magnetization switching of $TaIrTe_4/Fe_3GaTe_2$ under different bias fields parallel to the sample surface and current ($H_x$). **e.** The benchmark of SOT spin Hall conductivity vs. power consumption with state-of-the-art results[8,9,16,27,30,33–36]. Ellipse represents error and device to device variation in the calculated parameters.

We further investigated the impact of deterministic SOT switching on the external in-plane magnetic field parallel to the current direction (Fig. 4d). The external in-plane magnetic field ($H_x$) can break the symmetry of deterministic SOT switching. As the strength of $H_x$ increases, the switching mechanism transitions from being predominantly driven by the out-of-plane spin torque component ($\tau_{DL}^{z}$) to being influenced by the in-plane components ($\tau_{DL}^{x,y}$). We observed that a small positive $H_x$ has minimal effect on the SOT switching signal, however, increasing $H_x$ beyond 100 mT results in a noticeable reduction of the signal magnitude. Despite this reduction, the switching efficiency was maintained at 50%, demonstrating some robustness against the external magnetic field. In contrast, when $H_x$ is applied in the negative direction, the switching efficiency drops significantly to about 50% even at -10 mT, and it nearly diminishes to ~10% at -300 mT. Interestingly, the switching polarity remains unchanged up to 100 mT, indicating the effectiveness of the out-of-plane spin polarization of $TaIrTe_4$ in counteracting the external magnetic field[9,14]. In conventional SOT, where magnetization switching is driven purely by in-plane spin current, the switching polarity typically reverses abruptly with $H_x$[22,31]. However, this was not observed in our experiments, highlighting the larger contribution of $\tau_{DL}^{z}$ from $TaIrTe_4$ in the magnetization dynamics of $Fe_3GaTe_2$. In device 4 (data provided in Supplementary Fig. S2), we observed that the switching polarity remained unchanged even up to 200 mT when pulse current of ±4 mA was applied along the a-axis of $TaIrTe_4$. However, it abruptly reversed when both the current and magnetic field of similar magnitude was applied along the b-axis of $TaIrTe_4$.

Furthermore, to examine the presence of $\tau_{DL}^{z}$ and calculate unconventional SOT driven switching efficiency, we have performed AHE loop shift measurement with bias current (see Supplementary Fig. S3)[37,38]. The out-of-plane antidamping torque can shift the AHE hysteresis loop when a positive and negative dc bias current beyond a threshold value equivalent to switching current density is applied along the a-axis of $TaIrTe_4$. Such AHE loop shift ($H_{shift}$) is observed for compensating $\tau_{DL}^{z}$ driven intrinsic damping in $Fe_3GaTe_2$[9,14,37]. The SOT efficiency ($\varepsilon_{SOT}$) due to unconventional $\tau_{DL}^{z}$ torque is defined by equation[38–40]

$$\varepsilon_{SOT} = \frac{2eM_s\eta t_{FM}}{\hbar}\frac{H_{shift}}{J_{switch}} \quad \text{(Eq. 4)}$$

In our device, the $\varepsilon_{SOT}$ is 1.76, with the $H_{shift}$ and $J_{switch}$ calculated to be 2 mT and $1.81 \times 10^{10}\,\mathrm{Am^{-2}}$, respectively and $M_s$ taken as $0.97 \times 10^5$ $\mathrm{Am^{-1}}$ (see Supplementary Note 4). The switching efficiency parameter ($\eta$), defined as the ratio of switching current-driven and magnetic field-driven AHE, is observed to be 1 (Fig. 4c). Using the device parameters ($\varepsilon_{SOT} = 1.76$ and charge conductivity of $TaIrTe_4$ $\sigma_c$ =$9.4\times10^5$ S/m), we estimate the spin Hall conductivity in $TaIrTe_4/Fe_3GaTe_2$ heterostructure to be $\sigma_{SH}$ =ħ/2e($\varepsilon_{SOT}.\sigma c$) =$1.65\times10^6$ ħ/2e $(\Omega\mathrm{m})^{-1}$. By employing both SOT-induced magnetic switching and 2nd harmonic Hall measurements, we have established that the magnetization of $Fe_3GaTe_2$ in heterostructure with $TaIrTe_4$ can be effectively manipulated with a switching current density of $J_{switch}$~$1.81 \times 10^{10}\mathrm{A/m^2}$ and power density P ($= J_{switch}^2/\sigma_c$) of $0.348 \times 10^{15}\,\frac{\mathrm{W}}{\mathrm{m^3}}$ at room temperature. The benchmarked of the spin Hall conductivity $\sigma_{SH}$ and power density P of $TaIrTe_4/Fe_3GaTe_2$ devices along with literature available on state-of-the-art SOT devices [8,9,14,16–18,30,31,40,43,45-47] are shown in Fig. 4e and Supplementary Table 1.

**Calculation of unconventional spin Hall effect in $TaIrTe_4$**

Our experimental observations strongly suggest the presence of unconventional spin Hall effect in $TaIrTe_4$, which originates from the in-plane charge current and results in an out-of-plane spin-polarized spin current across the interface, corresponding the $\sigma^{Z}_{ZX}$ component of the spin Hall conductivity (SHC) tensor ($I^{Sz}_{Z} = \sigma^{Z}_{ZX} I^{C}_{X}$ where $I^{S}$ and $I^{C}$ are spin and charge currents, respectively). While similar effects were also found in another low-symmetry Weyl semimetal ($T_D$-$WTe_2$)[6,7,15,16,18], the symmetry constraints theoretically prohibit this configuration, and the experimental results have not been explained. Like $WTe_2$, $TaIrTe_4$ has low crystal symmetry described by space group (SG) 31 ($Pmn2_1$), consisting of a mirror plane perpendicular to the a axis (See Fig. 5a), as well as glide reflection and two-fold screw rotation, which prevents an unconventional SHC component[42].

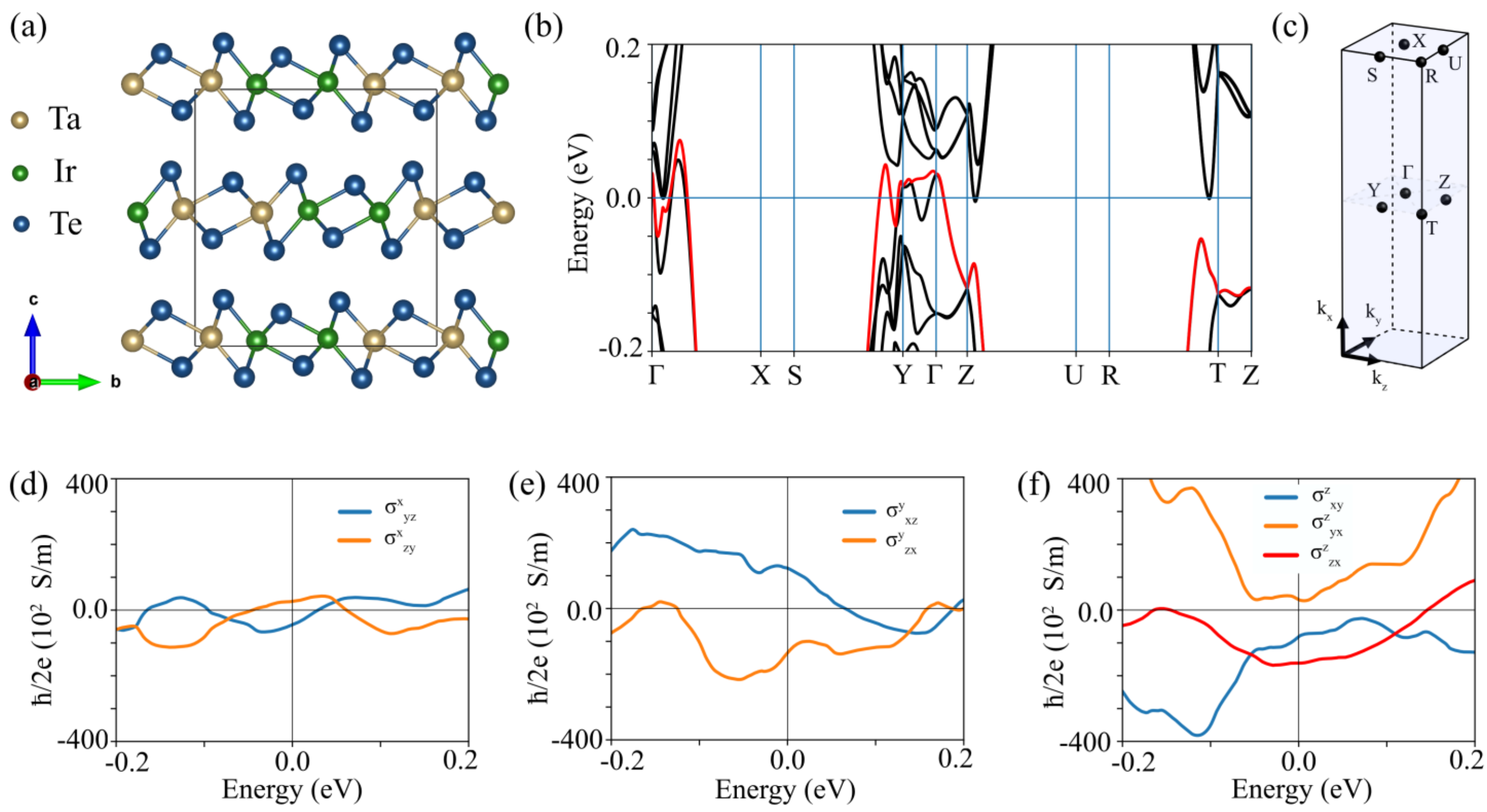

**Figure 6. Electronic structure and spin Hall conductivity calculations of $TaIrTe_4$ a.** Orthorhombic crystal unit cell of $TaIrTe_4$. **b.** Electronic structure of $TaIrTe_4$ calculated via density functional theory along the high-symmetry lines in the Brillouin zone shown in c. **d,e,f.** Calculated intrinsic spin Hall conductivity, representing configurations of spin current with spin polarization along a-, b-, and c-axis, respectively. The magnitude of the unconventional spin Hall conductivity $\sigma^{z}_{zx}$ at the Fermi level is determined mostly by the band marked as red in b.

To unveil the origin of the unconventional SHE, we have performed first-principles calculations of $TaIrTe_4$ (see Methods for computational details). As shown in Fig. 6, our results reveal a large spin splitting of bands near the Fermi level due to SOC, and the presence of seven spin Hall conductivity components: six conventional ($\sigma^{i}_{jk}$ with

i≠j≠k) and one unconventional component $\sigma_{ZX}^{Z}$, with the latter showing a magnitude comparable to the conventional components. The unconventional SHE at the Fermi level reaches $\sigma_{SH}$ =1.56x10$^{4}$ ħ/2e (Ωm)$^{-1}$, in agreement with experimental values reported for $TaIrTe_4$ (1.47-5.44)x10$^{4}$ ħ/2e (Ωm)$^{-1}$ [8,9,18], which vary depending on experimental conditions and sample characteristics. Although previous studies attributed its presence to the topological properties of the surface[8], our calculations show that a large unconventional SHE occurs even in the bulk.

We analyzed the crystal symmetry in more detail by directly applying symmetry operations, revealing that the relaxed structures exhibit a slight deviation from SG 31. This small structural distortion reduces the symmetry to either SG 6 (Pm) or SG 1 (P1), depending on the specified numerical precision (see Supplementary Note 7 for details). In both cases, the two-fold screw symmetry which normally prohibits unconventional SHE is absent, thus allowing for out-of-plane spin polarization of spin current. Structural distortion could further increase near the surface, potentially enhancing the generated spin accumulation. Therefore, from a symmetry perspective, the unconventional SHE component is justified, while its magnitude arises from the electronic properties, as discussed in Supplementary Materials.

Our experiments indicate even larger SHE values than previously reported and reveal an additional component, $\sigma_{ZY}^{Z}$, induced by charge current along the mirror planes. This component was absent in the previous studies and does not emerge in bulk calculation, suggesting a possible role of interfacial effects. The overall enhancement of SHE could arise from the proximity to the magnet, though it is possible that the spin Hall effect in $Fe_3GaTe_2$ could also contribute to unconventional SOT[13,41,43–50]. This highlights the unique behaviors of the $TaIrTe_4/Fe_3GaTe_2$ heterostructure and suggests new avenues for exploring SOT in low-symmetry materials.

**Summary**

In summary, we demonstrated the potential of $TaIrTe_4/Fe_3GaTe_2$ vdW heterostructures for generating a large and tunable nonlinear 2$^{nd}$ harmonic Hall effect, and energy-efficient deterministic field-free magnetization switching at room temperature. By leveraging the unique properties of the topological Weyl semimetal $TaIrTe_4$ and the magnetic $Fe_3GaTe_2$ with strong PMA, our findings reveal a large non-linear Hall effect, substantial unconventional out-of-plane damping-like torque and a remarkably low switching current density, outperforming conventional systems. To unveil the origin of unconventional charge-spin conversion phenomena in $TaIrTe_4$, detailed first-principles calculations were performed considering crystal symmetry and its impact on the energy-dependent electronic structure and spin Hall conductivity. Finally, we measured a substantial and tunable damping-like torque and observed deterministic field-free magnetization switching at a very low current density offering a promising route to energy-efficient and external field-free spintronic technologies.

## Methods

**Single crystal growth:** $TaIrTe_4$ single crystals were synthesized by evaporating tellurium from a Ta-Ir-Te melt, with the crystal growth conducted at 850 °C and Te condensation at 720 °C[51]. $Fe_3GaTe_2$ single crystals were grown via a self-flux method using Fe, Ga, and Te with 99.99% purity in the molar ratio of 1:1:2 in an evacuated and sealed quartz tube. The solid reactions took place for 24 hr at 1273 K, followed by cooling to 1153 K within 1 hr and slowly cooling down to 1053 K within 100 hr[10].

**Device fabrication:** The van der Waals heterostructure samples were prepared by mechanically exfoliating nanolayers of $TaIrTe_4$ and $Fe_3GaTe_2$ crystals on top of each other on a $SiO_2$/Si wafer using the Scotch tape method inside a glove box. The top sample surface was immediately capped with a 2 nm $Al_2O_3$ layer to protect from degradation with time. For the Devs1-3 nearly rectangular-shaped flakes were selected and the $TaIrTe_4/Fe_3GaTe_2$ heterostructures were patterned to Hall-bar geometry using electron-beam lithography (EBL) and Ti (15 nm)/Au (250 nm) contacts were prepared by EBL and electron beam evaporation. For Dev4, flakes are in arbitrary shapes, therefore dry physical etching by Ar ion milling was used to fabricate well-defined Hall-bar devices.

**Spin-orbit torque 2nd harmonic measurements:** Spin-orbit torque was characterized using an in-plane 2nd harmonic Hall lock-in measurement technique. The $R_{xy}^{1\omega}$ and $R_{xy}^{2\omega}$ for an a.c. current $I^{\omega}$ of 213.3 Hz were simultaneously measured while rotating the sample in the plane (azimuthal angle $\phi_B$) under an external field $\mu_0 H_{ext}$. The harmonic measurements were conducted using a Lock-in SR830 to detect the in-phase 1st and out-of-phase 2nd harmonic voltages. The 2nd harmonic measurements in the high magnetic field range were performed with a Quantum Design cryogen-free PPMS DynaCool system, interfaced with the SR830 to record the 1st and 2nd harmonic voltages. The 1st harmonic signal is detected by putting the voltmeter in phase with the oscillator, whereas the 2nd harmonic signal is out of phase with the source signal.

**Spin-orbit torque switching measurements** were conducted in a vacuum cryostat with a magnetic field strength of up to 0.7 T. Electronic measurements were carried out using a Keithley 6221 current source and a Keithley 2182A nanovoltmeter. To monitor the longitudinal and transverse Hall resistances, Keithley 2182A nanovoltmeters were employed. For SOT-induced magnetization switching, the Keithley 2182A nanovoltmeters were used to observe the Hall resistances responses, while a Keithley 6221 AC source applied a pulse current of 50 microsecond (µs) through the device, followed by a DC read current of magnitude 500 µA.

**Density functional theory calculations** for bulk $TaIrTe_4$ were performed using the Quantum Expresso package[52,53] by employing the Perdew, Burke, and Ernzerhof (PBE) generalized gradient approximation (GGA) for exchange-correlation functional[54]. We used fully relativistic pseudopotentials and expanded the electron wave functions in a plane-wave basis with the energy cutoff of 80 Ry. We adopted an orthorhombic unit cell with the experimental lattice constants a = 3.77 Å, b = 12.42 Å, and c = 13.18 Å[55]. The atomic positions were relaxed with the force and energy convergence thresholds set to $10^{-3}$ Ry/Bohr and $10^{-4}$ Ry, respectively. The Brillouin Zone (BZ) was sampled following the Monkhorst-Pack scheme with the k-grids of 20×8×8 and adopting a Gaussian smearing of $10^{-3}$ Ry. For the post-processing analysis, we used the python package PAOFLOW, which projects the ab initio wavefunctions onto pseudo-atomic orbital (PAO) basis to construct tight-binding Hamiltonians [56,57], further interpolated to a denser grid of 80×40×40. The charge-to-spin conversion response tensors were calculated using the approaches implemented in PAOFLOW and described in the previous works[58–60].

## Acknowledgments

Authors acknowledge funding from European Comission (EU) Graphene Flagship, European Innovation Council (EIC) project 2DSPIN-TECH (No. 101135853), 2D TECH VINNOVA competence center (No. 2019-00068), Wallenberg Initiative Materials Science for Sustainability (WISE) funded by the Knut and Alice Wallenberg Foundation, EU Graphene Flagship (Core 3, No. 881603), Swedish Research Council (VR) grant (No. 2021–04821, No. 2018-07046), FLAG-ERA project 2DSOTECH (VR No. 2021-05925) and MagicTune, Carl Tryggers foundation, Graphene Center, Chalmers-Max IV collaboration grant, VR Sweden-India collaboration grant, Areas of Advance (AoA) Nano, AoA Materials Science and AoA Energy programs at Chalmers University of Technology, Dutch Research Council (NWO grant OCENW.M.22.063). The fabrication of devices was performed at Nanofabrication laboratory MyFab at Chalmers University of Technology. The calculations were carried out on the Dutch national e-infrastructure with the support of SURF Cooperative (EINF-10786) and on the Hábrók high-performance computing cluster of the University of Groningen.

**Corresponding author**

Correspondence to Saroj P. Dash (saroj.dash@chalmers.se)

**Contributions**

L.P., S.P.D. conceived the idea and designed the experiments. L.P. fabricated and characterized the devices with support from B.Z., R.N., L.S., H.B. and P.R. The $TaIrTe_4$ single crystals were grown by A.A. and M.A.H, while G.Z., H.W., and H.C. grew the $Fe_3GaTe_2$ single crystals. V.L., K.T. and J.S. performed, analyzed and described the density functional theory calculations. J.S. supervised theoretical calculations. L.P. and S.P.D. analyzed and interpreted the experimental data and wrote the manuscript, with comments from all the authors. S.P.D. coordinated and supervised the project.

**Note:** After preparation of this manuscript, we came across reports on magnetization switching in $TaIrTe_4/Fe_3GaTe_2$ system[33,61]. However, spin dynamics experiments to understand the spin-orbit torque phenomena in all-2D vdW heterostructures are so far lacking. In our manuscript, in addition to energy-efficient magnetization switching, we report a detailed understanding of unconventional and tunable SOT magnetization dynamics using 2$^{nd}$ harmonic measurements in all-vdW heterostructures for the first time.